# Exact N-Envelope-Soliton Solutions of the Hirota Equation

**Jian-Jun SHU**

*School of Mechanical & Aerospace Engineering, Nanyang Technological University*
*50 Nanyang Avenue, Singapore 639798*

**Abstract**

We discuss some properties of the soliton equations of the type, $\frac{\partial u}{\partial t} = S[u,\bar{u}]$, where $S$ is a nonlinear operator differential in $x$, and present the additivity theorems of the class of the soliton equations. On using the theorems, we can construct a new soliton equation through two soliton equations with similar properties. Meanwhile, exact $N$-envelope-soliton solutions of the Hirota equation are derived through the trace method.

The trace method, which has been applied to the Korteweg-de Vries equation [1], Modified Korteweg-de Vries equation [2], Kadomtsev-Petviashvili equation [3], sine-Gordon equation [4,5] and Gz Tu equation [6], is useful for understanding these equations. The $N$-soliton solutions and some other results of these equations have been derived through the trace method.

The present paper deals with an application of the trace method to the nonlinear partial differential equation as follows:

$$\frac{\partial}{\partial t}u + L_x u = N_x(u,\bar{u}) \tag{1}$$

where

$$L_x u = \sum_{k=0}^{N_1} \alpha_k \frac{\partial^k}{\partial x^k} u$$

$$N_x(u,\bar{u}) = \sum_{k=1}^{N_2} \beta_k \prod_{m=0}^{N_k} \left(\frac{\partial^m}{\partial x^m} u\right)^{r_{m,k}} \left(\frac{\partial^m}{\partial x^m} \bar{u}\right)^{s_{m,k}}$$

where $\alpha_k$, $\beta_k$ are complex constants, $r_{m,k}$, $s_{m,k}$ are nonnegative integers, $r_k = \sum_{m=0}^{N_k} r_{m,k}$, $s_k = \sum_{m=0}^{N_k} s_{m,k}$, $r_1 = r_2 = \cdots = r_{N_2} = r$, $s_1 = s_2 = \cdots = s_{N_2} = s$, $d = r + s \geq 2$, and $r$, $s$ satisfy one of the relations:

(i) $s \geq 1$, $r = s+1$

(ii) $s = 0$, $r \geq 2$.

Substituting the formal series

$$u = u^{(1)} + u^{(d)} + \cdots + u^{((d-1)n+1)} + \cdots \tag{2}$$

into equation (1). We obtain a set of equations for $u^{((d-1)n+1)}$ $(n = 0,1,2,\cdots)$:

$$\left(\frac{\partial}{\partial t} + L_x\right) u^{((d-1)n+1)} = \sum_{l_1=1}^{N} \cdots \sum_{l_{(d-1)n+1}=1}^{N} C^{(n)}(P_{l_1}, \bar{P}_{l_2}, \cdots, \bar{P}_{l_{(d-1)n}}, P_{l_{(d-1)n+1}}) \phi_{l_1}^2 \bar{\phi}_{l_2}^2 \cdots \bar{\phi}_{l_{(d-1)n}}^2 \phi_{l_{(d-1)n+1}}^2 \tag{3}$$

$$\left(\frac{\partial}{\partial t} + L_x\right) u^{((r-1)n+1)} = \sum_{l_1=1}^{N} \cdots \sum_{l_{(r-1)n+1}=1}^{N} C^{(n)}(P_{l_1}, P_{l_2}, \cdots, P_{l_{(r-1)n}}, P_{l_{(r-1)n+1}}) \phi_{l_1}^2 \phi_{l_2}^2 \cdots \phi_{l_{(r-1)n}}^2 \phi_{l_{(r-1)n+1}}^2 \tag{4}$$

where the equations (3) and (4) correspond to (i) and (ii) respectively,

$$\phi_k(x,t) = A_k(0)\exp(P_k x - \Omega_k t), \quad \Omega_k = \frac{1}{2} L_p(2P_k), \quad L_p(x) = \sum_{k=0}^{N_1} \alpha_k x^k,$$

$A_k(0)$ and $P_k$ are complex constants ($k = 1,2,\cdots,N$), and $C^{(0)} = 0$.

We can obtain the solutions for equations (3) and (4) in the following form:

$$u^{((d-1)n+1)} = \sum_{l_1=1}^{N} \cdots \sum_{l_{(d-1)n+1}=1}^{N} \pi^{(n)} \phi_{l_1}^2 \bar{\phi}_{l_2}^2 \cdots \bar{\phi}_{l_{(d-1)n}}^2 \phi_{l_{(d-1)n+1}}^2 \tag{5}$$

$$u^{((r-1)n+1)} = \sum_{l_1=1}^{N} \cdots \sum_{l_{(r-1)n+1}=1}^{N} \pi^{(n)} \phi_{l_1}^2 \phi_{l_2}^2 \cdots \phi_{l_{(r-1)n}}^2 \phi_{l_{(r-1)n+1}}^2 \tag{6}$$

where $\pi^{(0)}=1$ and

$$\pi^{(n)}=C^{(n)}/[L_p(2P_{l_1}+2\overline{P}_{l_2}+\cdots+2\overline{P}_{l_{(d-1)n}}+2P_{l_{(d-1)n+1}})-L_p(2P_{l_1})-\overline{L}_p(2\overline{P}_{l_2})-\cdots-\overline{L}_p(2\overline{P}_{l_{(d-1)n}})-L_p(2P_{l_{(d-1)n+1}})] \quad (7)$$

or

$$\pi^{(n)}=C^{(n)}/[L_p(2P_{l_1}+2P_{l_2}+\cdots+2P_{l_{(r-1)n}}+2P_{l_{(r-1)n+1}})-L_p(2P_{l_1})-L_p(2P_{l_2})-\cdots-L_p(2P_{l_{(r-1)n}})-L_p(2P_{l_{(r-1)n+1}})]. \quad (8)$$

**Theorem 1** *Let*

$$\frac{\partial}{\partial t}u+L_x^{'}u=N_x^{'}(u,\bar{u}),\quad \frac{\partial}{\partial t}u+L_x^{''}u=N_x^{''}(u,\bar{u})$$

*be two arbitrary equations that are defined by equation (1). If* $r^{'}=r^{''}$, $s^{'}=s^{''}$, $\pi^{'(n)}=\pi^{''(n)}$ $(n=0,1,2,\cdots)$, *then, for equation*

$$\frac{\partial}{\partial t}u+L_x^{*}u=N_x^{*}(u,\bar{u})$$

*where* $L_x^{*}=aL_x^{'}+bL_x^{''}$, $N_x^{*}(u,\bar{u})=aN_x^{'}(u,\bar{u})+bN_x^{''}(u,\bar{u})$, $a$, $b$ *are two arbitrary real numbers, we have* $\pi^{*(n)}=\pi^{'(n)}=\pi^{''(n)}$ $(n=0,1,2,\cdots)$.

Proof. We consider the case (ii) by mathematical induction. Obviously $\pi^{*(0)}=\pi^{'(0)}=\pi^{''(0)}=1$. Assume $\pi^{*(n)}=\pi^{'(n)}=\pi^{''(n)}$ $(n=0,1,2,\cdots,k)$. When $n=k+1$, from equation (4), $C^{*(k+1)}=aC^{'(k+1)}+bC^{''(k+1)}$ and from equation (8)

$$\pi^{*(k+1)}=C^{*(k+1)}/[L_p^{*}(2\sum_{m=1}^{(r-1)n+1}P_{l_m})-\sum_{m=1}^{(r-1)n+1}L_p^{*}(2P_{l_m})]=[aC^{'(k+1)}+bC^{''(k+1)}]/[aL_p^{'}(2\sum_{m=1}^{(r-1)n+1}P_{l_m})+bL_p^{''}(2\sum_{m=1}^{(r-1)n+1}P_{l_m})-a\sum_{m=1}^{(r-1)n+1}L_p^{'}(2P_{l_m})-b\sum_{m=1}^{(r-1)n+1}L_p^{''}(2P_{l_m})]$$

$$=\pi^{'(k+1)}=\pi^{''(k+1)}$$

For the case (i), we can prove it in the same manner.

We introduce two $N\times N$ matrices $B$ and $D$ whose elements are given respectively by $B_{mn}=[1/(P_m+P_n)]\phi_m(x,t)\phi_n(x,t)$, $D_{mn}=[1/(P_m+\overline{P}_n)]\phi_m(x,t)\overline{\phi}_n(x,t)$.

**Theorem 2** *Let*

$$\frac{\partial}{\partial t}u+L_x^{'}u=N_x^{'}(u,\bar{u}),\quad \frac{\partial}{\partial t}u+L_x^{''}u=N_x^{''}(u,\bar{u})$$

*be two arbitrary equations that are defined by equation (1). If they have respective solutions*

$$u^{'}=T_r[B_x^{'}f(D^{'}\overline{D}^{'})]\ (or\ T_r[B_x^{'}g(B^{'})]),\quad u^{''}=T_r[B_x^{''}f(D^{''}\overline{D}^{''})]\ (or\ T_r[B_x^{''}g(B^{''})])$$

*where* $f$, $g$ *are arbitrarily derivable functions in the neighbourhood of zero, then, for equation*

$$\frac{\partial}{\partial t}u+L_x^{*}u=N_x^{*}(u,\bar{u})$$

*where* $L_x^{*}=aL_x^{'}+bL_x^{''}$, $N_x^{*}(u,\bar{u})=aN_x^{'}(u,\bar{u})+bN_x^{''}(u,\bar{u})$, $a$, $b$ *are two arbitrary real numbers, we have solution* $u^{*}=T_r[B_x^{*}f(D^{*}\overline{D}^{*})]\ (or\ T_r[B_x^{*}g(B^{*})])$.

Proof. Since $f$, $g$ are arbitrarily derivable functions in the neighbourhood of zero, $f$, $g$ can expand into power series in convergence region. Corresponding, $u^{'}$, $u^{''}$ can expand into power series. Comparing the coefficients, we have $r^{'}=r^{''}$, $s^{'}=s^{''}$, $\pi^{'(n)}=\pi^{''(n)}$ $(n=0,1,2,\cdots)$. From Theorem 1, we obtain $u^{*}=T_r[B_x^{*}f(D^{*}\overline{D}^{*})]\ (or\ T_r[B_x^{*}g(B^{*})])$.

On using the Theorems 1 and 2, we can construct a new soliton equation through two soliton equations with similar properties. As an example, we use the trace method to solve the Hirota equation [7] as follows:

$$i\psi_t+i3\alpha|\psi|^2\psi_x+\rho\psi_{xx}+i\sigma\psi_{xxx}+\delta|\psi|^2\psi=0 \quad (9)$$

where $\alpha$, $\rho$, $\sigma$ and $\delta$ are positive real constants with the relation $\frac{\alpha}{\sigma}=\frac{\delta}{\rho}=\lambda$. In one limit of $\alpha=\sigma=0$, the equation reduces to the nonlinear Schrödinger equation [8] that describes a plane self-focusing and one-dimensional self-modulation of waves in nonlinear dispersive media.

$$i\psi_t+\rho\psi_{xx}+\delta|\psi|^2\psi=0. \quad (10)$$

In another limit of $\rho=\delta=0$, the equation for real $\psi$, reduces to the modified Korteweg-de Vries equation [9,10]

$$\psi_t+3\alpha|\psi|^2\psi_x+\sigma\psi_{xxx}=0. \quad (11)$$

Hence, the present solutions reveal the close relation between classical solitons and envelope-solitons. Substituting the formal series

$$\psi = \psi^{(1)} + \psi^{(3)} + \cdots + \psi^{(2n+1)} + \cdots \tag{12}$$

into equation (9), we obtain a set of equations for $\psi^{(2k+1)}$ $(k = 0,1,2,\cdots)$ :

$$i\psi_t^{(1)} + \rho\psi_{xx}^{(1)} + i\sigma\psi_{xxx}^{(1)} = 0 \tag{13}$$

$$i\psi_t^{(3)} + \rho\psi_{xx}^{(3)} + i\sigma\psi_{xxx}^{(3)} = -i3\alpha\psi^{(1)}\overline{\psi}^{(1)}\psi_x^{(1)} - \delta\psi^{(1)}\overline{\psi}^{(1)}\psi^{(1)} \tag{14}$$

$$\vdots$$

$$i\psi_t^{(2n+1)} + \rho\psi_{xx}^{(2n+1)} + i\sigma\psi_{xxx}^{(2n+1)} = -i3\alpha\sum_{l=0}^{n-1}\sum_{m=0}^{n-l-1}\psi^{(2l+1)}\overline{\psi}^{(2m+1)}\psi_x^{(2n-2l-2m-1)} - \delta\sum_{l=0}^{n-1}\sum_{m=0}^{n-l-1}\psi^{(2l+1)}\overline{\psi}^{(2m+1)}\psi^{(2n-2l-2m-1)} \tag{15}$$

$$\vdots .$$

We can solve the set of equations iteratively:

$$\psi^{(1)} = \sum_{l_1=1}^{N}\phi_{l_1}^2(x,t) \tag{16}$$

$$\psi^{(3)} = -\frac{\lambda}{8}\sum_{l_1=1}^{N}\sum_{l_2=1}^{N}\sum_{l_3=1}^{N}\frac{1}{(P_{l_1} + \overline{P}_{l_2})(\overline{P}_{l_2} + P_{l_3})}\phi_{l_1}^2(x,t)\overline{\phi}_{l_2}^2(x,t)\phi_{l_3}^2(x,t) \tag{17}$$

where $\phi_k(x,t) = A_k(0)\exp(P_k x - \Omega_k t)$ , $\Omega_k = -2i\rho P_k^2 + 4\sigma P_k^3$ , $A_k(0)$ and $P_k$ are complex constants relating respectively to the amplitude and phase of the $k$ th soliton $(k = 1,2,\cdots,N)$. We introduce two $N \times N$ matrices $B$ and $D$ whose elements are given respectively by

$$B_{mn} = [1/(P_m + P_n)]\phi_m(x,t)\phi_n(x,t), \quad D_{mn} = [1/(P_m + \overline{P}_n)]\phi_m(x,t)\overline{\phi}_n(x,t) .$$

With the matrices $B$ and $D$ , $\psi^{(1)}$ and $\psi^{(3)}$ are expressed as

$$\psi^{(1)} = T_r[B_x] \tag{18}$$

$$\psi^{(3)} = -\frac{\lambda}{8}T_r[B_x(D\overline{D})] . \tag{19}$$

In general, we can prove that

$$\psi^{(2n+1)} = (-1)^n\frac{\lambda^n}{8^n}T_r[B_x(D\overline{D})^n] , \quad n = 0,1,2,\cdots \tag{20}$$

satisfies the equation (15).

With the definitions of the matrices $B$ and $D$ ,

$$\psi^{(2n+1)} = (-1)^n\frac{\lambda^n}{8^n}\sum_{1}\cdots\sum_{2n+1}\frac{\phi_1^2\overline{\phi}_2^2\cdots\overline{\phi}_{2n}^2\phi_{2n+1}^2}{(P_1 + \overline{P}_2)(\overline{P}_2 + P_3)\cdots(P_{2n-1} + \overline{P}_{2n})(\overline{P}_{2n} + P_{2n+1})} . \tag{21}$$

Here and in the following we simplify the expressions by writing $1,2,\cdots,2n+1$ instead of $l_1,l_2,\cdots,l_{2n+1}$. There should be no confusion about this. We have

$$i\psi_t^{(2n+1)} + \rho\psi_{xx}^{(2n+1)} + i\sigma\psi_{xxx}^{(2n+1)} = (-1)^n\frac{\lambda^n}{2^{3n-1}}\sum_{1}\cdots\sum_{2n+1}\{4i\sigma[(P_1 + \overline{P}_2 + \cdots + \overline{P}_{2n} + P_{2n+1})^3 - (P_1^3 + \overline{P}_2^3 + \cdots + \overline{P}_{2n}^3 + P_{2n+1}^3)]$$

$$+ 2\rho[(P_1 + \overline{P}_2 + \cdots + \overline{P}_{2n} + P_{2n+1})^2 - (P_1^2 - \overline{P}_2^2 + \cdots - \overline{P}_{2n} + P_{2n+1}^2)]\}\frac{\phi_1^2\overline{\phi}_2^2\cdots\overline{\phi}_{2n}^2\phi_{2n+1}^2}{(P_1 + \overline{P}_2)(\overline{P}_2 + P_3)\cdots(P_{2n-1} + \overline{P}_{2n})(\overline{P}_{2n} + P_{2n+1})} . \tag{22}$$

Substituting two identities

$$(k_1 + k_2 + \cdots + k_{2n} + k_{2n+1})^3 - (k_1^3 + k_2^3 + \cdots + k_{2n}^3 + k_{2n+1}^3) = 3\sum_{l=0}^{n-1}\sum_{m=0}^{n-l-1}[(k_1 + \cdots + k_{2l+1})(k_{2l+1} + k_{2l+2})(k_{2l+2m+2} + k_{2l+2m+3})$$

$$+(k_{2l+1} + k_{2l+2})(k_{2l+2m+2} + k_{2l+2m+3})(k_{2l+2m+3} + \cdots + k_{2n+1})] \tag{23}$$

$$(k_1 + k_2 + \cdots + k_{2n} + k_{2n+1})^2 - (k_1^2 - k_2^2 + \cdots - k_{2n}^2 + k_{2n+1}^2) = 2\sum_{l=0}^{n-1}\sum_{m=0}^{n-l-1}[(k_{21+1} + k_{2l+2})(k_{2l+2m+2} + k_{2l+2m+3})] \tag{24}$$

into equation (22) and using (20) for $\psi^{(2k+1)}$ $(k < n)$ , we obtain

$$i\psi_t^{(2n+1)} + \rho\psi_{xx}^{(2n+1)} + i\sigma\psi_{xxx}^{(2n+1)} = -\delta\sum_{l=0}^{n-1}\sum_{m=0}^{n-l-1}\psi^{(2l+1)}\overline{\psi}^{(2m+1)}\psi^{(2n-2l-2m-1)} - i\frac{3}{2}\alpha\sum_{l=0}^{n-1}\sum_{m=0}^{n-l-1}\overline{\psi}^{(2m+1)}(\psi^{(2l+1)}\psi^{(2n-2l-2m-1)})_x .$$

Therefore we obtain the $N$ -envelope-soliton solution for equation (9) in the following form:

$$\psi = T_r\{\sum_{k=0}^{+\infty}(-1)^k\frac{\lambda^k}{8^k}[B_x(D\overline{D})^k]\} = T_r[B_x(1 + \frac{\lambda}{8}D\overline{D})^{-1}] \tag{25}$$

where $\left\|D\overline{D}\right\| < \frac{8}{\lambda}$ in a certain region. In particular, for $N = 1$, we obtain the one-envelope-soliton solution:

$$\psi(x,t) = \frac{A_1(0)}{2}\mathrm{sech}[(P_1 + \overline{P}_1)x - (\Omega_1 + \overline{\Omega}_1)t + \eta]\exp[(P_1 - \overline{P}_1)x - (\Omega_1 - \overline{\Omega}_1)t - \eta] \tag{26}$$

where $\eta = \frac{1}{2}\ln\left(\frac{\lambda|A_1(0)|^4}{8(P_1 + \overline{P}_1)^2}\right)$.